\documentclass[twocolumn,preprintnumbers,superscriptaddress]{revtex4}
\usepackage{epsf}
\usepackage{graphicx}
\usepackage{dcolumn}
\usepackage{bm}
\def\prb{Phys. Rev. B}
\def\prl{Phys. Rev. Lett.}

\def\be{\begin{equation}}
\def\ee{\end{equation}}
\def\ba{\begin{eqnarray}}
\def\ea{\end{eqnarray}}

\begin{document}

\draft

\title{Drag resistance of 2D electronic
microemulsions}

\author{Boris Spivak}
\email[]{spivak@dirac.phys.washington.edu}
\affiliation{Department of Physics, University of Washington, Seattle, WA 98195}
\author{Steven A. Kivelson}
\email[]{stevek@physics.ucla.edu}
\affiliation{Department of Physics, University of California, Los
  Angeles, California  90095}

\begin{abstract}
Motivated by recent experiments of Pillarisetty {\it et al}, \prl
{\bf 90}, 226801 (2003), we present a theory of drag in electronic
double layers at low electron concentration. We show that the drag effect
in  such systems is anomolously large, it has unusual temperature and
magnetic field dependences accociated with the Pomeranchuk effect, and
does not vanish at zero temperature.

\end{abstract}


\maketitle

Over the last decade a large number of dramatic phenomena have
been discovered in studies of the two dimensional electron gas
(2DEG) in MOSFET \cite{Abrahams,Myriam} and semiconductor heterojunction
devices \cite{Shahar,Alan,Pillarisetty}. Collectively, these
phenomena are referred to under the euphonious title of the ``2D
apparent metal insulator transition.'' They are associated with
the behavior of the 2DEG in very clean (high mobility) devices
with strong interactions, {\it i.e.} when the ratio of the typical
interaction strength to the Fermi energy, $r_s= 1/\sqrt{na_B^2}$,
is large, $r_s \sim 10 - 40$.  (Here $n$ is the electron density
per unit area and $a_B$ is the effective Bohr radius.)  These
observations make it clear that the physics of this simple, but
conceptually vital system is much richer than was previously
appreciated. Elsewhere, we have proposed
\cite{SpivakPS,SpivakKivelsonPS} that these phenomena are
associated with a set of ``electronic microemulsion phases'' which
occur as a consequence of Coulomb frustrated phase
separation\cite{physicaC} in the ideal (zero disorder) 2DEG at
densities intermediate between the high densities, where the
system is a Fermi liquid, and low densities, where it is a Wigner
crystal.

In the present paper we concentrate on the drag resistance of 2D
large $r_s$ metals, with the goal of explaining the unexpectedly
large value of the drag resistance and the unusual temperature and
magnetic field dependances that have recently been reported in
bilayer p-type GaAs heterojunction devices
\cite{Pillarisetty,Pillarisetty1}. In a bilayer device, in the
presence of a dissipative current $I_{A}$ in the first (active)
layer, the inter-layer electron-electron interactions induce a
momentum transfer between layers, which in turn produces a
 voltage, $V_{P}$, across the second (passive) layer. In
the linear regime this effect can be characterized by a drag
resistance (per square)
\begin{equation}
R_{D}=V_{P}/I_{A} .
\label{drag}
\end{equation}
Measurements of $R_D$ are a potentially useful probe of
interesting correlations of an electron fluid. It is exceedingly
small at low temperatures in a Fermi liquid while, as we shall
see, it is parametrically larger in a ``bubble fluid'' phase.  It
is therefore an ideal diagnostic for the presence of such phases.

 The existing
theories of drag \cite{Price,MacDonnald}  are mostly based on
the Fermi liquid theory of the electron liquid in individual
layers. A common feature  is that, at
 small temperatures, $R_D$ is small and vanishes as $T\to 0$
or as the spacing between layers, $d\to \infty$:
\begin{equation}
R_{D}\sim (h/e^2)(k_Fd)^{-\alpha_d}(T/E_F)^{\alpha_T}.
\label{alpha}
\end{equation}
For a Fermi liquid, $\alpha_T=2$ up to  logarithmic corrections
and $\alpha_d=2-4$ depending on the value of $k_F\ell$,
where $\ell$ is the elastic mean-free path and $k_F$ is the Fermi
momentum.

{\bf  Experimental Background:}
 The drag experiments that motivated this study are performed on
high mobility samples whose conductances are significantly larger
than $e^{2}/\hbar$.  Therefore, the putative electron localization
length one would infer on the basis of localization theory is
exponentially long, and hence the associated 2D localization
phenomena can be ignored for present purposes.

 In double layer devices with relatively
small $r_s$, measurements\cite{Gramila}
of the drag are in
qualitative agreement with Fermi liquid theory
\cite{Price,MacDonnald}.
 However, at large $r_{s}$,   experiments on
p-GaAs double layers \cite{Pillarisetty,Pillarisetty1} differ
 significantly from the predictions of this
theory :

{\bf 1.} The drag resistance in these samples found to be 2-3
orders of magnitude larger than expected on the basis of Fermi
liquid theory.

{\bf 2.} In the presence of a magnetic field, $H_{\|}$, parallel to the film,
 the temperature dependence of $R_{D}(T)$ is significantly
suppressed. (Here, $H_{\|}$ primarily couples to the spins of the
electrons, rather than to their orbital motion. )
  Whereas in a Fermi liquid, $R_{D}(T)$ is a
quadratic function of $T$, in large $r_s$ devices, at $H_{\|}=0$,
the value of $\alpha_T$ in Eq. \ref{alpha}
 appears to be noticeably larger than 2;  for example, $\alpha_{T}=2.7$ in
 \cite{Pillarisetty,Pillarisetty1}.
The value of
$\alpha_T(H_{\|})$ then decreases
 with $H_{\|}$
 and saturates for $H_{\|}>H^{*}$ at a value which is significantly smaller
than 2;
 for example, in Refs. \cite{Pillarisetty} and  \cite{Pillarisetty1},
$\alpha_{T}(H_{\|}>H^{*})\sim
 1.2$.  $H^*$ is generally believed to be the field required to fully polarize
the electron spins.

{\bf 3.}   As a function of increasing $H_{\|}$, $R_{D}(H_{\|})$
increases by a factor of 10-20  and saturates when $H_{\|}>H^{*}$
(See Fig.1 in \cite{Pillarisetty} ). In the framework of the Fermi
liquid theory, one would expect a significant {\it decrease} of
$R_{D}(H_{\|})$ in the spin polarized state  because $E_F$ is
increased by a factor of 2 as $H_{\|}$ increases from 0 to
$H_{\|}>H^{*}$, thus decreasing the electron-electron scattering
rate.

{\bf 4.} The $T$ and especially the $H_{\|}$ dependences of
$R_{D}(T, H_{\|})$ and the resistances of the individual layers
$R(H_{\|}, T)$ look qualitatively similar to one another. (See
Figs.1 a,b in \cite{Pillarisetty} )

In fact, $R(H_{\|}, T)$, itself, appears highly anomalous
from the viewpoint of Fermi liquid theory.  Thus, although our
primary focus will be on $R_D$, we will also summarize the most
salient anomalies in $R(H_{\|}, T)$. Since similar phenomena  have
been observed in studies of the 2DEG in MOSFETs and of the 2D hole
gas (2DHG) in semiconductor heterojunction devices, we report
observations from both systems:

{\bf 5.} An apparent continuous metal insulator transition as a
function of $n$  has been observed in Si MOSFET's \cite{Abrahams}
and to a lesser extend in GaAs heterojunctions
\cite{Shahar,Alan,Pillarisetty}. (No such transition is predicted
on the basis of Fermi liquid theory.) For the most part, in this
paper we will focus on the slightly higher density samples on the
metallic side of the apparent transition.

{\bf 6.} The resistance of these samples increases significantly
with increasing $T$:  by as much as a factor of 6 in Si MOSFET's
\cite{Abrahams} and as much as 3  in GaAs
heterojunctions\cite{Alan,Pillarisetty}.

{\bf 7.} The resistance of metallic samples at low $T$ is
significantly increased by non-zero $H_{\|}$, and becomes nearly
$H{\|}$ independent at high magnetic fields, $H_{\|}>H^{*}$
\cite{Abrahams}. Moreover, non-zero $H_{\|}$ strongly suppresses
the $T$ dependence of $R$, so much so that $R$ is nearly
temperature independent for $H_{\|} > H^*$. (To date, this latter
effect has been documented only in Si MOSFET's \cite{Myriam},
where $dR/dT$ at low temperature is decreased for $H_{\|} > H^*$
by as much as a factor of 100 relative to $H_{\|}=0$.)

{\bf  Theoretical Background:}  The phase diagram of the bilayer
2DEG depends on  many parameters. For simplicity, in this article
we consider only the case when the distance between the layers,
$d$, is larger than the inter-electron distance so that the
discrete nature of the electrons is irrelevant in the calculation
of the interaction energy between electrons in different layers
\cite{reza}.  Moreover, we restrict ourselves to the case in which
the ``passive'' layer has a large electron density, $n_P$ ({\it
i.e.} small $r_s$), while the ``active'' layer has a low electron
density, $n_A$ ({\it i.e.} $r_s\sim r_c$, where $r_c$ is the
critical value of $r_s$ for the liquid-crystal transition --  for a
single band with an isotropic effective mass, $r_c$ is
estimated\cite{Ceperley} to be $r_c\approx 38$.)  In this limit,
the passive layer acts as a ground-plane, so the sequence of
microemulsion phases in the active layer is the same as was
derived previously\cite{SpivakPS,SpivakKivelsonPS} for the 2DEG in
a MOSFET.

For zero quenched disorder, the FL to WC transition is always
expected to be first order \cite{brazovskii}. In the absence of
long-range interactions, this would lead to a regime of density
exhibiting two-phase coexistence; due to the screening by
electrons in the passive layer, the electrons in the active layer
interact via a dipolar interaction. It has been shown in
\cite{SpivakPS} that in this case, first order transitions are
forbidden because  the energy of a long interface between the two
phases is negative.  As a result, there exists between the WC (for
$n_A < n^{c}_{WC}$) and the FL (for $n_A
> n^{c}_{FL}$), an intermediate range ($n^{c}_{WC} < n_A < n^{c}_{FL}$)
of densities  in which there is a set of new ``microemulsion''
phases which on the mean field level can be characterized as a
mixture of microphase separated regions of liquid and crystal
\cite{SpivakPS,SpivakKivelsonPS}. The relative concentration,
sizes, shapes, and organization of these regions is determined by
thermodynamics. Even the full enumeration and classification of
the distinct phases that result from Coulomb frustrated phase
separation is not complete, much less a thorough investigation of
their properties.

 To be concrete, we
will focus on the drag effect in the case when the active layer is
in the bubble phase in which the majority phase is a FL, with a
finite concentration of bubbles of WC.  This phase can be viewed
as a suspension of pieces of Wigner crystallites floating in an
otherwise uniform Fermi fluid, like ice in a river. (Many of the
same considerations apply to more general micro-emulsion phases,
as we will discuss in a forthcoming
paper\cite{RezaKivelsonSpivak}.) As a function of decreasing,
$n_A$, the areal fraction of Wigner crystal, $f_{WC}$ rises
continuously from vanishingly small when $n_A=n^{c}_{FL}$ to
$f_{WC}=1$ when $n_A=n^{c}_{WC}$.

The WC bubble phase occurs when $n_{A}$ is close to $n^{c}_{FL}$
(See Figs.1 and 2 in \cite{SpivakPS} and \cite{SpivakKivelsonPS}). As $n_A\to
n^{c}_{FL}$,  the bubble size approaches a constant value $L_{B} =
d\exp[\gamma]$,  where $\gamma$ is a (positive and order 1)
dimensionless ratio of microscopic parameters which is
proportional to the surface tension between the Wigner crystal and
Fermi fluid phases. The spacing between bubbles
$[n^{c}_{FL}-n_A]^{-1/2}$ diverges as $n_A \to n^{c}_{FL}$. At
mean-field level, these bubbles, themselves, form a crystal of
bubbles, but for sufficiently dilute bubbles, either quantum and
classical fluctuations inevitably melt the bubble crystal.
Nevertheless the charge inhomogeneities in the resulting bubble
liquid are still very slowly fluctuating compared to the
relaxation times of the Fermi liquid. The competition between the
local tendency to phase separation and the Coulomb interaction
(capacitive energy), results in a  mean density difference between
the WC and FL regions $\Delta n \sim \sqrt{n_A}/d$.

The qualitative  temperature and magnetic field dependence of
$f_{WC}$ is determined by the Pomeranchuk effect
\cite{SpivakPS,SpivakKivelsonPS}. Because the Wigner crystal has
higher spin entropy density ($S_{WC} \sim n \log[2]$) than the
Fermi liquid ($S\sim n (T/E_F)$), $f_{WC}$ is an increasing
function of temperature at low temperatures.  This is a precise
analogue of the Pomeranchuk effect in $He^{3}$. Similarly, since
the spins are substantially more polarizable in the Wigner crystal
phase, $f_{WC}$ is an increasing function of $H_{\|}$, but it
saturates above a (temperature dependent) characteristic magnetic
field strength, $H^*$ at which the spins are fully polarized.
Moreover, the two effects compete, so that when $H_{\|} \gg T$,
the temperature dependence of $f_{WC}$ is quenched. (Here, and
henceforth, we will adopt units in which $\hbar=k_B=$ the Bohr
magneton $=1$.)

The strong dependence of $f_{WC}$ on $n_A$, $T$ and $H_{\|}$ is
the origin of the strong dependences of all other properties of
the system on these physical variables.  To get a feeling for the
expected dependences, consider the case in which $T$ and $H_{\|}$
are small enough that the Fermi liquid free energy can be well
approximated by its zero temperature value, but large enough that
we can ignore the (exponentially small\cite{roger,sudip} in powers
of $\sqrt{r_s}$) magnetic exchange energy in the Wigner crystal.
In this case, and in the absence of disorder, the difference in
free energy per unit area between the uniform WC and FL phases is
\be \Delta F(n_A,T,H_{\|}) = \Delta E(n_A) - n_A
T\log[2\cosh(H_{\|}/2T)]. \label{DeltaF} \ee Since $f_{WC}$ is
determined by the competition between the local tendency to phase
separation and the Coulomb interaction, it is a smooth function of
$\Delta F$; we can obtain the $T$ dependence of $f_{WC}$ from the
$T$ dependence of $\Delta F$ by the chain rule.  For $H_{\|}=0$,
since $\Delta F$ is a linear function of $T$, it follows that
$f_{WC}$ is as well.

{\bf Drag in a dilute bubble liquid:}  The reasons that $R_D$ is
so small in a Fermi liquid are not hard to fathom.  The
fluctuations in the charge density in the active layer are small
when $T/E_F$ is small, and are further reduced when averaged over
a large length scale, $d\gg 1/\sqrt{n_A}$, hence the dependances
exhibited in Eq. \ref{alpha}.  In a microemulsion WC bubble phase,
by contrast, there are large amplitude charge inhomogeneities  on
the length scale, $L_B$, which implies strong coupling to the
electrons in the passive layer.

In the regime when the bubble concentration is small there are two
types of the current carriers in the active layer: electrons and
 crystalline bubbles. At zero temperature the electron
Fermi-liquid contribution to the drag resistance vanishes and the
drag effect is entirely due to motion of Wigner crystallites.

Each Wigner crystal bubble in the active layer casts an image
potential in the passive layer.  As bubbles in the active layer
move with respect to the passive one, the electrons in the passive
layer scatter on the moving image potential.  To compute the drag
resistance, we compute the mean force per unit area, $F_{PA}$,
exerted on the electron fluid in the passive layer when a current
$J$ is passed through the active layer.  In the linear response
regime, it is clear that $F_{PA} \propto n_B v_B$ where $n_B =
f_{WC}/\pi L_{B}^2$ is the concentration of bubbles and $v_B$ is
the mean drift velocity of the bubbles. Thus
 \be
 R_D =  \frac 1
{e^2}\ A_D \ \mu_B \ \frac {n_{B}} {n_A } \label{RD}
\ee
 where  $\mu_B \equiv v_B/\bar v$ is the relative mobility of the
 bubbles, $\bar v \equiv J/en_A$ is the mean electron drift velocity in
the active layer, and $A_D$ is a dimensionless constant (discussed
below) which depends on the properties of the electron liquid in
the passive layer.

In the absence of disorder, $\mu_B\to 1$.  Moreover, as we will
discuss below, $\mu_B$ remains  non-zero in the presence of weak
disorder so long as we are dealing with a bubble liquid phase.
Thus,  $R_D$ is similarly non vanishing as $T\to 0$! Moreover, at
low temperatures, the dominant $T$ and $H_{\|}$ dependances of
$R_D$ are through the dependence of $f_{WC}$ on these variables -
{\it i.e.} the Pomeranchuk effect discussed above.

The physics of the proportionality constant $A_D$ is in general,
complicated - similar complexities occur as in the closely related
problem of electro-migration in metals. To compute $A_D$ we need
to assess the nature of the scattering potential induced in the
passive layer by a bubble in the active layer.  The long range
tails of potential are screened away by the electrons in the
passive layer.  The magnitude of the potential, $V_B \sim ed\Delta
n$, is determined
 by the magnitude
 of the charge deficit in the active layer, $\Delta n$,
associated with the bubble.   From the previously mentioned
scaling of $\Delta n$ with $d$, it follows that  the WKB
barrier penetration factor, $W=\sqrt{meV_B} L_{B}$ is larger than
1, so the potential behaves like a hard object of radius $\sim L_{B}$
being dragged through the passive layer - electron tunnelling
through the image potential is negligible. This is one of the key
reasons that a bubble phase produces a large drag resistance.

It requires additional microscopic analysis to obtain an estimate
of $A_D$, and there are numerous ``regimes'' depending on the
relative values of $n_B^{-1/2}$ (the mean spacing between
bubbles), $ L_{B}$, and the elastic and inelastic mean-free paths,
$\ell_{el}$ and $\ell_{e-e}$, in the passive layer.  The important
general point is that $A_D$ is not generally small.

To give a concrete example, consider the case in which $n_B^{-1/2}
\gg \ell_{el}\gg  L_{B}$ and $\ell_{ee} \gg \ell_{el}$. In this
case, each bubble acts as an uncorrelated scattering center with
cross section proportional to $ L_{B}$, and consequently \be A_D =
L_{B}n_{P}^{1/2} \ee In a future paper, we will report an analysis
of $A_D$ in the various other regimes.

{\bf Bubble mobility.}  In the absence of quenched disorder, there
are two distinct bubble phases - the bubble crystal and the bubble
fluid.

In the bubble liquid in the absence of pinning, the bubbles are
swept along with the fluid, so that $\mu_B= 1$ and $R=R_D$.
  In the presence of weak disorder, we still
expect that $\mu_B\approx 1$ at temperatures large compared to the
pinning potential, although in this case, the additional channels for momentum
relaxation in the active layer will generally result in $R > R_D$.

The $T\to 0$ behavior of $\mu_B$ is a more subtle issue.  At
$n_{A}d^{2}\gg 1$ the bubbles are large compared to the spacing
between electrons.  To first order they can be treated as
classical objects - at this level of approximation, they are
pinned at favorable sites of the disorder potential. Quantum
mechanically, however, the bubbles can tunnel from one pinning
site to another.  Moreover, the fact that the number of bubbles is
only conserved on average affects the nature of this tunnelling
process profoundly. Specifically, since the Fermi liquid degrees
of freedom are ``fast,'' we can integrate them out to obtain an
effective action for the bubbles.  (This can be done simply, using
the method of Levitov and Shytov\cite{LevitovShytov}, as we will
show in a forthcoming paper.)  Among other interactions that are
generated by doing this is an effective bubble hopping matrix
element, $t(\vec R,\vec R')\sim |\vec R -\vec R'|^{-\beta}$ where
$\beta \propto 1/G$ and $G$ is the conductance of the Fermi fluid
in units of $e^2$. Physically, this derives from a process in
which a bubble at pinning site $\vec R$ virtually melts, the
associated electron density deficit propagates through the Fermi
fluid, and then recrystallizes at site $\vec R'$.   It is well
known \cite{Levitov} that for $\beta < D$, where $D=2$ is the
spatial dimension, Anderson localization does not occur. This
establishes that, for clean enough systems, the bubbles, although
large, are not localized.  $\mu_B$ is typically a fairly strongly
increasing function of $T$, but it approaches a non-zero value as
$T\to 0$ if $G$ is large enough!

{\bf Qualitative comparison with experiment:} The theoretical
model sketched above contains the ingredients necessary for a
qualitative understanding of much concerning the listed
experiments:

The drag resistance in the WC bubble phase $R_{D}$ is not
parametrically small, either in powers of $1/k_F d$ or of $T/E_F$,
as in the Fermi liquid, which accounts for point {\bf 1} in the
above. According to Eq. 5  $R_D$ is proportional to $n_B$ and
$\mu_B$. Since $L_{0}$ is $T$ and $H_{\|}$ independent,
$n_B\propto f_{WC}$, so the general features of the $T$ and
$H_{\|}$ dependances of $R_D$ and $R$, points {\bf 2}, {\bf 3},
{\bf 4}, {\bf 6}, and {\bf 7}, follow directly from the
Pomeranchuk effect, and the interplay between the $T$ and $H_{\|}$
dependence is readily understood from Eq. \ref{DeltaF}. In
particular, $f_{WC}$ and, consequently, $R_{D}$ are increasing
function of $H_{\|}$ which saturate at $H_{\|}>H^{*}$.  When the
spins are polarized by $H_{\|}
> T$, they no longer have any entropy, so the $T$ dependence of
$f_{WC}$ and the corresponding contribution to the $T$-dependence
of $R_{D}(T)$ are quenched.

In the case of very small disorder $\mu_{B}$ is close to unity and
does not exhibit strong $T$ and $H_{\|}$ dependances. Then $R(T,
H_{\|})$ is entirely determined by the aforementioned dependances
of $n_{B}$. The more detailed $T$-dependances of $R_{D}(T,
H_{\|})$, points {\bf 2} , cannot be discussed without a
microscopic calculation of $\mu_{B}(T, H_{\|})$. .

Since the dominant $T$ and $H_{\|}$ dependances of $R_D$ and $R$
derive from their implicit dependence of $f_{WC}$, it is
unsurprising that these dependances are similar.

It is an unavoidable consequence of the existence of microemulsion
phases that, in the absence of quenched disorder, there is a
sequence of {\it continuous} phase transitions (as opposed to a
putative first order WC to FL transition) between different
microemulsion phases and between these phases and the low and high
density uniform phases. These phases become increasingly less
conducting with decreasing $n_A$, until the limiting WC phase is
insulating.  Thus, for weak enough disorder, an apparent metal
insulator transition is inevitable, thus accounting for the
observation that gave the subject its name, point {\bf 6}.
Since $\mu_{B}$ does not vanish as $T\to 0$ the conceptually key
conclusion of our analysis is  that $R_D$ does not vanish as $T\to
0$ in the WC bubble phase: $R_{D}(T=0)\neq 0$. This fact has not
been confirmed experimentally, perhaps because $R_{D}(T=0)$ is
small.

Finally we would like to compare our approach with that of Ref.
\cite{Sarma}, to the experimental data on the drag resistance. The
approach of Ref. \cite{Sarma} is supposed to work well at high
electron concentration far from the regime of phase separation.
Extending the results of \cite{Sarma} to the regime of low $n$
one, in principle, can explain the large value of $R_{D}$. On the
other hand we suspect that the dramatic increase of
$R_{D}(H_{\|})$ in the magnetic field and quenching it's
temperature dependence in low $n$ samples can not be simply
explained using that approach.

\acknowledgements

 This work was supported in part by the National Science
Foundation under Contracts No. DMR-01-10329 (SAK) and DMR-0228104
(BS).  We thank L. Levitov, R. Pillarisetty, B. Shklovskii, D.
Tsui, and S.Das Sarma  for useful discussions.


\begin{thebibliography}{99}
\bibitem{reza} We have
recently shown \cite{RezaKivelsonSpivak} that the scaling of
$L_{B}$ with $d$ is only valid so long as $n_{A}$ is sufficiently
close to the critical concentration $n^{c}_{FL}$. At smaller
$n_{A}$ the value of $L_{D}$ can be significantly smaller than
$d$, and even of order $n_{A}^{-2}$.
\end{thebibliography}
\end{document}